\documentclass[
  journal=pasa,
  manuscript=research-paper, 
  year=202X,
  volume=Y,
]{cup-journal}
\usepackage{microtype,siunitx,booktabs}
\usepackage{amsmath, amssymb}
\usepackage{subcaption}
\usepackage{float}

\usepackage[inkscapearea=page]{svg}
\usepackage{adjustbox}
\usepackage{xcolor}
\usepackage{soul}
\title{Increasing the detectability of long-period and nulling pulsars in next-generation pulsar surveys}

\author{G.~Grover}
\affiliation{International Centre for Radio Astronomy Research, Curtin University, Bentley, WA 6102, Australia}
\email[G. Grover]{garvit.grover@icrar.org}

\author{N.~D.~R.~Bhat}
\affiliation{International Centre for Radio Astronomy Research, Curtin University, Bentley, WA 6102, Australia}

\author{S.~J.~McSweeney}
\affiliation{International Centre for Radio Astronomy Research, Curtin University, Bentley, WA 6102, Australia}

\doi{ZZZ}

\received {dd Mmm YYYY}
\revised  {dd Mmm YYYY}
\accepted {dd Mmm YYYY}
\published{dd Mmm YYYY}


\newcommand{\dmu}{cm$^{-3}$\,pc}
\newcommand{\psrtwo}{J0026-1955 }
\newcommand{\psrfive}{J0452-3418 }
\newcommand{\psrone}{J0036-1033}
\newcommand{\psrthree}{J1002-2036}
\newcommand{\psrfour}{J1357-2530}
\newcommand{\snpp}{S/N$_{\rm{PP}}$}

\begin{document}

\begin{abstract}

Recent discoveries of multiple long-period pulsars (periods ${\sim}10\,$s or larger) are starting to challenge the conventional notion that coherent radio emission cannot be produced by objects that are below the 
{many theorised death lines.}
Many of the past pulsar surveys and software have been prone to selection effects that restricted their sensitivities towards long-period and sporadically-emitting objects. Pulsar surveys using new-generation low-frequency facilities are starting to employ longer 
{dwell times, which makes them significantly more sensitive}
to detecting long-period or nulling pulsars. There have also been software advancements to aid more sensitive searches towards long-period objects. Furthermore, recent discoveries suggest that nulling may be a key aspect of the long-period pulsar population. We simulate both long-period and nulling pulsar signals, using the Southern-sky MWA Rapid Two-meter (SMART) survey data as reference, and explore the detection efficacy of popular search methods such as the fast Fourier transform (FFT), fast-folding algorithm (FFA) and single pulse search (SPS). For FFT-based search and SPS, we make use of the PRESTO implementation,  and for FFA we use RIPTIDE. We find RIPTIDE's FFA to be more sensitive; however, it is also the slowest algorithm. PRESTO's FFT, although faster than others, also shows some unexpected inaccuracies in detection properties. SPS is highly sensitive to long-period and nulling signals,
{but only for pulses with high intrinsic signal-to-noise ratios.}  
We use these findings to inform current and future pulsar surveys that aim to uncover a large population of long-period or nulling objects and comment on how to make optimal use of these methods in unison.


\end{abstract}

\section{INTRODUCTION }
\label{sec:int}

Coherent radio emission from pulsars is thought to rely critically on pair production within their magnetospheres \citep{1975Ruderman}. At longer rotation periods ($P$), the potential difference in the magnetosphere is suspected to be insufficient for pair production to occur, and hence, no emission can be produced \citep{1975Ruderman, 1979ApJ...231..854A}. This naturally places a constraint on the rotation period $P$ and the characteristic magnetic field $B \propto \sqrt{P\dot{P}}$ of the pulsar, where $\dot{P}$ is the period 
derivative. The related constraints can be identified as `death lines' on a $P$-$\dot{P}$ diagram. Many death lines have been proposed, each with different assumptions such as the magnetic field orientation, assumptions about the pair production, etc \citep[eg][]{1975Ruderman, Chen1993, Zhang2000, rea2023longperiod}.
However, the one that most completely encompassed the pulsar population was by \cite{Zhang2000}. 

\begin{figure}
    \centering
    \includegraphics[width=\linewidth]{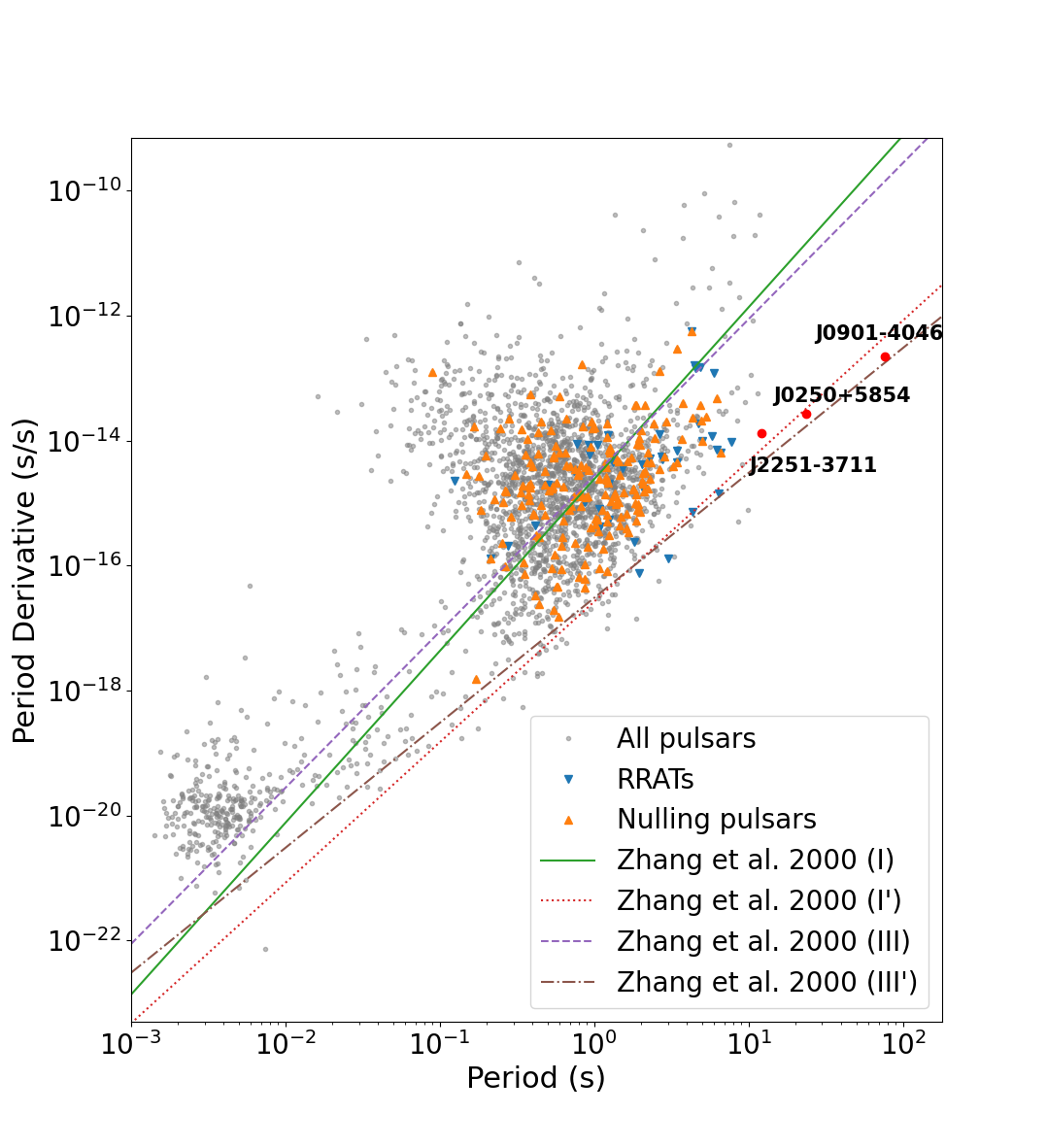}
    \caption{A $P\dot{P}$ diagram showing all pulsars (grey), nulling pulsars (orange) and RRATs (blue). Also highlighted are the three recent long-period pulsar discoveries {\citep{Tan2018,Morello2020a, Caleb2022}}. The various death lines, taken from \cite{Zhang2000}, are also shown, as per the legend. These death lines are improvements based on the works of \cite{1975Ruderman} and \citep{Chen1993}.}
    \label{fig:ppdot}
\end{figure}

In recent years, there has been a flurry of discoveries of pulsars with very long periods that challenge some of the most conservative death line models, as shown in Figure \ref{fig:ppdot}. These include a 12.1-second pulsar J2251-3711 discovered by \cite{Morello2020a}, a 23.5-second pulsar J0250+5854 by  \cite{Tan2018} and PSR J0901-4046 with a rotation period of 76 seconds \citep{Caleb2022}. These are currently among the slowest known radio pulsars. Along with their relatively small $\dot{P}$, their positions on the $P$-$\dot{P}$ diagram are also 
near the edge of the furthest (least stringent) death line. This death line is shown as \cite{Zhang2000} III' in Figure \ref{fig:ppdot}. The discoveries of such pulsars also suggest the possible existence of a larger population of long-period pulsars that could be uncovered 
using more optimal search strategies. 

Remarkably, all three  objects show some degree  of ``nulling,'' which is the cessation of detectable radio emission for a certain duration or number of rotations. For example, PSRs J2251-3711 and J0250+5854 have nulling fractions of 67\% and 27\% respectively \citep{Tan2018, Morello2020a}, and PSR J0901-4046 shows nulling occurring on time scales shorter than a single pulse
\citep{Caleb2022}. Assuming these objects are representative of the population, nulling may be a common property among these objects. It has been suggested that nulling is indicative of the radio emission becoming unsustainable for a pulsar and the nulling fraction increases progressively until there is no detectable emission from the pulsar \citep{1976MNRAS.176..249R, 2007MNRAS.374.1103Z}. However, no 
convincing correlation has been found between nulling and physical parameters such as $P$ or $\dot{P}$ 
\citep{Anumarlapudi2023, Sheikh2021}, though this may be also due to observational selection effects. 

In addition to these three pulsars, a couple of pulsar-like sources with extremely long periods have also been detected in recent years. A noteworthy source is GLEAMX J162759.5-523504.3,  with a rotation period of $\sim 18.18$\,min, discovered by \citet{2022Natur.601..526H}. 
Another similar source, with an even longer period of $\sim$21 min, has just been discovered \citep{Hurley-Walker2023}.  The origins of this new class of sources remain currently unknown. The first source does seem to go into a quiescent state, which could be interpreted as an extended nulling state, but interestingly the second source has been active for over three decades.  
These discoveries have motivated \cite{rea2023longperiod} to extend the original models for death lines.

Yet another recent  addition is 
a  $\sim10$\,s pulsar discovered by  \citet{Su2023} 
using the Five-hundred-metre Aperture Spherical Telescope (FAST).
Intriguingly, over a 2.29-year observational time span, there is no evidence of nulling in this object. Regardless, it  sits well beyond the current death line models on the $P$-$\dot{P}$ diagram, thereby further strengthening the evidence of a possible long-period pulsar population. Uncovering a larger population of sources (both pulsars or pulsar-like objects) can help significantly advance our understanding of coherent radio emission in pulsars and current models for the death lines. 

Death line models are typically modified to constrain new long-period pulsars as they are discovered; however, long-period pulsar discoveries are made difficult due to observational selection effects.
Most large pulsar surveys that scan large swaths of the skies typically have dwell times between 2\,min and 5\,min  and therefore are less sensitive to long-period pulsar signals \citep{2014ApJ...791...67S, 2021RAA....21..107H, 2010MNRAS.409..619K}. Longer dwell times naturally increase the  sensitivity to these objects, particularly those with long durations of nulling or rotation periods larger than tens of seconds. Specifically, longer observing duration increases the chance of detecting a larger number of  pulses, especially when $P \sim 10 - 100 $\,s or longer. The emergence of wide-field instruments at low frequencies such as the Murchison Widefield Array (MWA; \citealt{2013PASA...30....7T}) and Low Frequency Array (LOFAR; \citealt{2013A&A...556A...2V}) and their adaptation to large pulsar surveys \citep[e.g.,][]{Bhat2023smart1, 2019A&A...626A.104S}, means longer dwell times are now affordable, at least at low frequencies. This increases the achievable sensitivity in general, particularly for those objects with rotation periods $\gtrsim$ 10\,s, or with long null durations ($\sim$tens of minutes). The LOFAR Tied-Array All-Sky Survey (LOTAAS; \citet{2019A&A...626A.104S}) has a 60\,min dwell time, and the Southern-sky MWA Rapid Two-meter (SMART) survey uses an 80\,min dwell time. The SMART survey uses the MWA to observe the whole southern sky between 140\,MHz - 170\,MHz. The data has 100\,$\mu$s time resolution and 10\,kHz frequency resolution, and are stored as raw voltages, and can be beamformed offline 
to tessellate the full field of view using thousands of sensitive tied-array (phased array) beams. 
For more 
details on 
the SMART survey see \citet{Bhat2023smart1}. In detecting a long-period or highly intermittent population of pulsars, these surveys will have a competitive advantage.

\subsection{Overview of Search Methods}

    {In addition to the use of more optimal survey strategies, it is also important to adopt search algorithms that are sensitive to sporadic, intermittent and or long-period objects. In principle, long-duration data from surveys like the SMART can also be exploited to search for pulsars in binary systems; however, our focus will primarily be on isolated or single pulsars.}
There are three popular pulsar searching methods, including the fast Fourier transform (FFT) based search, the fast folding algorithm (FFA; first described by \citealt{Staelin1969}) and the single pulse search (SPS). Fourier methods have been the primary ones used in pulsar searching and have been very successful, having discovered the vast majority of the pulsars currently known. However, in searching for long-period pulsars, Fourier-based methods have been shown to be inherently less sensitive towards long-period and narrow duty-cycle signals, which is generally the case for most long-period pulsars \citep{2017MNRAS.467.1661V, Cameron2017, Singh2022, 2009ApJ...702..692K, Morello2020, Parent2018}. A major obstacle is the presence of low-frequency `red' noise that arises from slowly varying power levels in the system (e.g. due to gain variations), or radio frequency interference (RFI) that is persistent in nature.
This tends to mask long-period signals in Fourier space. Moreover, this method relies on the signal to be periodic to make a significant detection and is therefore inherently less sensitive towards the sporadic emission from pulsars that tend to null for long durations. PRESTO \citep{2001PhDT.......123R}, includes algorithms for suppressing red noise, and has not been robustly tested for its efficacy in detecting such objects that emit sporadic or intermittent signals.

FFA is a time-domain periodicity search. It searches for pulsars by efficiently folding signals at multiple trial periods. There have been a few different implementations of this algorithm: ffaGo \citep{Parent2018}, ffancy \citep{Cameron2017} and RIPTIDE \citep{Morello2020}, all highlighting the advantages of using FFA over FFT-based searches, especially for long-period pulsars. The algorithm has been demonstrated in its ability to find new pulsars, having recently detected 
{6 new pulsars}
in the GMRT High-Resolution Southern-sky (GHRSS) survey \citep{Singh2023}. FFA has also been shown to be effective at finding long-period pulsars, and in fact, has been able to successfully make independent detections of PSRs J0250+5854 and J2251-3711, at a higher detection significance than FFT methods \citep{Morello2020a, Tan2018}. 

Comparisons between FFA and FFT have also been explored in \citealt{Singh2022} (hereafter S22). They made use of both simulations and examples of real data and investigated the efficacies of the FFT and FFA implementations in order to compare the quality of detections. They varied the period, duty cycle and morphology of pulses and {reported FFA to be able to detect pulsar signals at a consistently higher S/N than FFT.}

Similar to FFT, FFA is less effective at detecting sporadic signals. However, a robust analysis comparing the efficacies of FFA's implementations in detecting nulling signals has not been carried out yet. Some limited studies have been published on this front; e.g. the recent work by \citet{Singh2023IV} who compare FFA and FFT in detecting nulling signals. However, this was done for only one particular source.

Finally, single pulse search (SPS) is quite effective in detecting pulsed, transient-like emission, and is routinely used in transient, pulsar and fast radio burst (FRB) surveys. It searches for emission by convolving the time series with box car filters with various trial widths (e.g. in logarithmic steps of $2^n$, where $n = 0, 1, 2, ..., n$).
This method is demonstrably effective at finding long-period pulsars, having successfully made the initial detections of both PSRs J2251-3711 and  \citep{Tan2018} J0901-4046 \citep{Caleb2022}. SPS 
has also been effective in detecting sporadically emitting pulsars, such as Rotating Radio Transients (RRATs), which are, by definition, better detected by SPS \citep{2011BASI...39..333K}. Although the efficacy of SPS has been demonstrated in detecting both 
sporadic and long-period sources,  the efficacy of SPS towards sporadic and long-period objects has not been robustly tested and compared with other algorithms.

Ongoing large surveys, such as the SMART survey, intend to incorporate FFA and SPS into their search pipelines, alongside FFT-based methods, in order to increase sensitivity towards long-period objects \citep{Bhat2023smart2}. Given that recent long-period pulsar discoveries tend to show a significant amount of nulling, it is valuable to test the efficacies of various methods in detecting such long-period nulling pulsars. In order to test these methods, we chose PRESTO's implementations of FFT and SPS, and RIPTIDE's implementation of FFA (see justifications in Section \ref{sec:sim_signals}). We perform a simulation analysis to explore the large search parameter space, covering periods, nulling parameters and S/Ns for a robust analysis. Section \ref{sec:meth} describes the process and the justifications for our analysis. For reference, we try to duplicate typical observations from the SMART survey. The results are presented in Section \ref{sec:simresults}. As a sanity check, we also explore the effectiveness of the search methods in detecting pulsars in real data  (Section \ref{sec:realdata_results}). Finally, we discuss our results and make some recommendations to consider in future search processing 
to increase the detectability of long-period pulsars (Section \ref{sec:dis}). Our conclusions are summarised in Section \ref{sec:con}.

\section{Analysis}
\label{sec:meth}

    

\begin{figure}
    \centering
    \includegraphics[width=\textwidth]{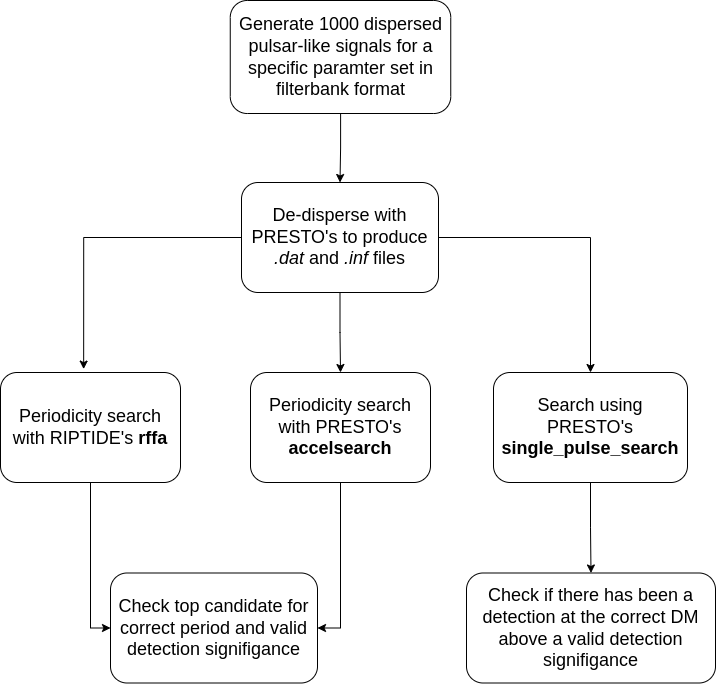}
    \caption{A flow diagram for the simulation of the synthetic pulsar signal and their processing through FFT, FFA and SPS methods.}
    \label{fig:flow_diagram}
\end{figure}

In order to test how well FFT, FFA and SPS would perform in detecting a variety of long-period and nulling pulsars, we generated a large number of simulated pulsar signals. This allows us to sample a large parameter space, uninfluenced by selection biases. A high-level overview of our simulation and searching procedure is shown in Figure \ref{fig:flow_diagram}.

S22 performed a similar study where they tested the detection significance of RIPTIDE's FFA and PRESTO's FFT implementations using simulated data. We chose to keep parts of our analysis similar to S22 to allow for meaningful comparison. S22 varied periodicity, subpulse morphology and duty cycles. As we are interested in nulling properties of long-period pulsars, we chose to vary the pulse period, nulling fraction, nulling duration and signal strength, searching with RIPTIDE's FFA, PRESTO's FFT and PRESTO's SPS implementations.

\subsection{Simulating pulsar signals}\label{sec:sim_signals}

The simulated signals consisted of trains of Gaussian pulses (mimicking pulsar signals) injected into Gaussian noise (receiver noise). We based our simulations on the SMART survey, as it has the longest dwell time of any current survey (4800\,s), and therefore, in principle, offers better detection prospects (within the sensitivity limits) for detecting long-period signals.  

Our main goal is to explore the long-period and nulling parameter spaces. We, therefore, simulate signals with varying periods ($P$), nulling properties (nulling fraction, $n_{\rm f}$; nulling duration, $N_{\rm d}$) and average signal-to-noise ratio per pulse (\snpp). Within each simulated signal, we allowed individual pulses to vary in brightness according to a modulation index that was kept constant for all simulations. We also kept the pulse duty cycle constant at $4\%$. 

The target parameters were varied logarithmically with 6 - 7 increments in order to encapsulate a large range of values. The periods ranged from 0.5\,s - 50\,s. The choice of a large value for the maximum period was driven by our intent to simulate very long-period pulsars ($P \gtrsim 10$\,s).

To emulate nulling, we varied $n_{\rm f}$ from 1\% - 90\% and $N_{\rm d}$ from 5 pulses - 1000 pulses. This accounts for the majority of the nulling pulsar population as well as extremely nulling pulsars such as RRATs. For each null sequence within a simulation, its duration was drawn from a Poissonian distribution with an expected value $N_{\rm d}$. Burst sequence durations were drawn from a Poissonian distribution with expected value:
\begin{equation}
    B_{\rm d} = \frac{N_{\rm d}}{n_{\rm f}} - N_{\rm d}.
    \label{eq: burstd}
\end{equation}
These choices guarantee that the long-term behaviour of the simulated pulse train matches the expected behaviour for a pulsar with the selected average $N_{\rm d}$ and $n_{\rm f}$, whilst still reflecting the randomness of nulling behaviours observed in real pulsars.

Finally, as pulsars are known to vary in brightness, we chose S/N per pulse (\snpp) to range from 0.1 - 10. A large upper limit was chosen primarily because at low frequencies ($\lesssim 300$\,MHz), pulsars tend to be generally brighter and some RRATs tend to have bright pulses \citep{2010MNRAS.402..855B}. Given that the number of pulses within a simulated signal was not predetermined due to the randomness of the nulls, the S/N was chosen per pulse rather than the signal as a whole. Varying \snpp{} also allows us to more accurately test the threshold of SPS capabilities than we would have with an integrated S/N. The distribution of \snpp{} was assumed to be log-normal, in agreement with what is commonly observed \citep[e.g.][]{2012MNRAS.423.1351B}:

\begin{equation}
    f(x) = \frac{1}{x \sigma \sqrt{2\pi}} \exp \left( \frac{(\ln x - \mu)^2}{2\sigma^2} \right).
\end{equation}
Here $\mu$ is the mean \snpp{} (where \snpp{} ranges from 0.1 - 10) and $\sigma^2$ is the variance, which is dependent on the modulation index $m$ of the signal via $\sigma = m\mu$. The modulation index was chosen arbitrarily as 0.1 
{as the value can range from 0 - 2 for typical and sub-pulse drifting pulsars \citep{2006A&A...445..243W}.}


Due to computational constraints, all signals were generated with a single dispersion measure  (DM) of 150\,\dmu. 
{As we chose not include any scattering effects, the choice of DM was arbitrary as it has little to no effect on the search algorithms. If scattering was implemented to the simulated data, a large DM such 150 \dmu{} would imply considerable scatter broadening, which would lower the peak amplitude of the pulse. The effects of scattering appear differently in the time and frequency domain and hence would affect FFT, FFA and SPS differently. As our primary focus is to assess the effects of nulling on various search methods, we have not considered effects such as pulse broadening (scattering) that alter the pulse shape (and thus impact the detectability).}
As there is no scattering and the data are dedispersed before being searched, the DM is only relevant as metadata. As long as the algorithm is allowed to search signals of the given DM, it has no effect on the search process. In practice, however, the exact DM of the pulsar may not be searched, and there would be some loss of S/N, as described by \cite{Cordes2003}, in the case of searching sporadic emission. This would directly affect the detection significance of the search methods but since we are using simulated data, we take the `ideal case' approach.

\subsection{Searching Methods on Simulated Data}\label{sec: searchmethods}

Once generated, these signals were processed and searched with RIPTIDE's FFA, PRESTO's FFT and PRESTO's SPS. We chose RIPTIDE as it is the latest, progressive implementation of FFA which has parallelised computation and improved detectability with matched filtering (see \cite{Morello2020} for details). This is also the implementation used in S22.

For the FFT implementation, we chose PRESTO's \textbf{accelsearch} as it is the most commonly used version and is already in use in the first pass (i.e. shallow survey) of the SMART survey \citep{Bhat2023smart1}. Although \textbf{accelsearch} was designed to do an additional acceleration search in order to target binary pulsars, we do not use this feature as our simulated signals do not include any Doppler modulation of the period due to binary orbital motion. By using the PRESTO implementation for comparison with other searches in this work, we could directly relate our analysis with S22.

The SPS method was originally described in \cite{Cordes2003}. More recent implementations 
\citep[e.g.][]{2018MNRAS.480.3457M}
primarily improve additional aspects such as the machine learning classification of candidates. We chose PRESTO's \textbf{single\_pulse\_search} as it is essentially an implementation of the original algorithm described in \cite{Cordes2003}. 

It is important to acknowledge that FFA, FFT and SPS are inherently different searching methods that analyse different aspects of the signal and report their results using different metrics. For example, PRESTO's \textbf{accelsearch} searches in Fourier space and reports a $\sigma_{\rm FFT}$ (the detection significance) and, through additional code, a spectral S/N (S/N$_{\rm FFT}$). RIPTIDE's FFA also reports a detection period but it is accompanied by a S/N$_{\rm FFA}$ measured in the time domain. PRESTO's \textbf{single\_pulse\_search} on the other hand, is a transient search method, that reports a $\sigma_{\rm SPS}$ for its transient candidate. The $\sigma$ and S/N for these methods are thus not equivalent though they all signify detection significance. Hence, a more meaningful comparison of these different methods involves testing whether or not the method can detect a signal based on their relevant criteria. 

\subsubsection{FFA \& FFT}

For both FFA and FFT searches, detections would be successful if a candidate was produced with the correct period, within some tolerance, and with a detection significance $\geq$ 6$\sigma$. As FFA does not report a $\sigma$ value, we used the reported S/N$_{\rm FFA}$. 
{A S/N$_{\rm FFA}$ cut-off of $\sim$8 was recently chosen by \citep{2024MNRAS.527.3208W} and the $\sigma_{\rm FFT}$ cut-off can be $2\sigma_{\rm FFT} - 6\sigma_{\rm FFT}$ \citep{2019A&A...626A.104S,2014ApJ...791...67S} for surveys, depending on the instrument.} It was also acceptable if the detection was a fractional harmonic of the fundamental period (up to the 32nd harmonic) while still being within the tolerance. The 32nd harmonic limit was chosen as the FFT algorithm was allowed to sum up to 32 harmonics to search for periodic signals. We also lowered the default low-frequency cutoff from one to 0.5 Hz to accommodate long-period signals. The period tolerance was defined as
\begin{equation}
    \text{Tol} = \frac{P^2\times D}{T}\times1.2
\end{equation}
where $D$ is the duty cycle of the signal (which was set to $\sim$ 4\%) and $T$ is the total signal time (4800\,s). This was formulated such that the difference in the detected period and original period should not be larger than the full-width half max of the pulse. This ensured that signals folded at the detected period retained a high S/N and resembled the original signal. {The additional factor of 1.2 was enforced as the tolerance was too strict and did not reflect a realistic tolerance for surveys, which are more flexible for the detection period.}

RIPTIDE was our chosen implementation for FFA. It comes with a search pipeline, \textbf{rffa}, which uses all the tools within the package to search data for periodic signals. This pipeline must be accompanied by a configuration file that specifies the search parameters, such as DM, S/N$_{\rm FFA}$ cut-off, running median width, period range, etc.

\subsubsection{SPS}

SPS is not a periodicity search; for a given candidate, \textbf{single\_pulse\_search} (PRESTO) primarily returns a DM and detection significance, $\sigma_{\rm SPS}$ and S/N$_{\rm SPS}$. 
{The $\sigma_{\rm SPS}$ is calculated as $\rm S/N_{\rm raw}/\rm \sqrt{bin\_width}$ and is reported in a text file, and S/N$_{\rm SPS}$ is reported in a diagnostic plot; however, the two values are identical, implying that the plot may have intended to display $\sigma_{\rm SPS}$ but it was incorrectly labelled, or vice versa. Hence in this paper, we will only report $\sigma_{\rm SPS}$. Note that the bins mentioned here refer to time bins of the convolved time series generated by \textbf{single\_pulse\_search}.}
As mentioned earlier, all our signals were generated with an arbitrary DM of 150\,\dmu{}. If SPS produced a candidate above a significance of $6\sigma_{\rm SPS}$, that would be considered a detection. 
{This cutoff is similar to the cut-off of 5$\sigma_{\rm SPS}$ used by \citep{2019A&A...626A.104S}. For FFA and FFT, the S/N cut-off of 6 was based on other surveys, however, for SPS, the cut-off was chosen based on trials with false candidates. We intentionally de-dispersed the signals at various incorrect DMs and found all false candidates at $\sigma_{\rm SPS} < 6$, motivating the choice of $\sigma_{\rm SPS} = 6$ as our threshold for real detections.}

Realistically, pulsar data are trialled at multiple DMs as a real pulsar signal retains its signal strength over multiple DMs, as described by \cite{Cordes2003}. Detections of a transient at multiple DMs help confirm the validity of a candidate. However, the aforementioned approach would only assist the detection quality, if the quality was assessed by a machine learning algorithm or by visual inspection. In this work, we are testing the algorithm's ability to find simulated signals, hence the quality of detection is irrelevant.

\section{RESULTS FROM SIMULATED DATA}\label{sec:simresults}

We simulated pulsar signals with varying periods, nulling fractions, nulling durations and S/N per pulse. Each combination of parameters had 1000 signals generated. These were searched using PRESTO's \textbf{accelsearch} (FFT), RIPTIDE's \textbf{rffa} (FFA) and PRESTO's \textbf{single\_pulse\_search} (SPS).

The results from our simulations are summarised in Figures \ref{fig:pvnf} \& \ref{fig:pnd}. They are shown as heatmaps with the colour representing the percentage of simulated signals detected out of 1000, as indicated by the colour bar. Each figure has 7 heatmap panels with a progressively increasing \snpp{}. Each of the 7 plots shows the period of the signal vs. either the nulling fraction (Figure \ref{fig:pvnf}) or nulling duration (Figure \ref{fig:pnd}). In either case, the nulling parameter that is not plotted has been averaged in order to make the data presentable. For example, Figure \ref{fig:pvnf} plots the nulling fraction and rotation period, hence the information relating to the nulling duration has been averaged.

In general, all search methods share some common characteristics. The number of detections increases for all methods as the \snpp{} increases. There are also more detections made for shorter periods. This is reconcilable as signals with shorter periods and brighter pulses tend to be present more frequently, even at high values of nulling parameters (i.e. $n_{\rm f}$ and $N_{\rm d}$). 

When comparing nulling duration and period, a slight decrease in the number of detections is seen between FFA and FFT. This may be due to our approach of drawing the nulling duration values from a Poisson distribution. At low mean values {($\sim 5$)}, the Poisson distribution has a low chance of outputting 0. Based on equation \ref{eq: burstd}, if the null duration $N_{\rm d}$ is 0 pulses, the burst duration $B_{\rm d}$ would also be 0 pulses. Effectively, this would result in an additional null sequence, deviating the mean null duration to a larger value. This is a possible reason why the number of detections is lower for $N_{\rm d} = 5$ pulses than for consecutive higher nulling durations.

\subsection{Benchmarking}

PRESTO's \textbf{accelsearch} was the fastest process and took typically $\lesssim 0.5$\,s to search a signal {summing up to 8 harmonics; however, when summing up to 32 harmonics searching took up to 2-3\,s on some occasions}. The \textbf{accelsearch} routine was used with mostly default settings, searching for signals with 0 acceleration,{ summing up to 32 harmonics and a lower frequency minimum of 0.5\,Hz to accommodate for the long-period signals}. PRESTO's \textbf{single\_pulse\_search} was the second fastest, computing for $\lesssim 1$\,s for low \snpp{} {($<0.5$)} signals and 1\,s - 2\,s for high \snpp{} (0.5 - 1) signals. This was also used with the default option, { meaning the width of boxcar filters used were $\leq 30$\,bins; if the whole range of widths were used (1 - 300 bins), the processing time would likely be longer}. Finally, RIPTIDE's \textbf{rffa} was the slowest implementation, which could range {3\,s - 20\,s} depending on the period ranges searched and the definition of other parameter ranges in the configuration file. {Note these process times were from a local computer and not a supercomputer, hence computing times may change with devices and different algorithm implementations. \footnote{{Local machine has an AMD Ryzen 6 core/12 thread CPU and 2x32GB RAM}}}

\subsection{FFT}


{Overall \textbf{accelsearch} performed well in detecting long-period and nulling signals, contrary to common perception. The performance was comparable to that of FFA's; however, FFT had noticeably fewer detections at low \snpp{} (0.1 - 0.5) and at high nulling fractions ($n_{\rm f}$ = 0.5 - 0.9). FFT's better performance is likely due to the very long observation length (4800\,s), which can greatly increase the sensitivity to long-period and nulling signals. We did not consider effects such as RFI and red noise, which can heavily limit the detection of long-period signals in the Fourier space \citep{Parent2018, Cameron2017, 2017MNRAS.467.1661V}. The duty-cycle was also kept constant at 4\%, rather than decreasing with increasing period; a narrow duty-cycle has been shown to be difficult to detect by \textbf{accelsearch} \citep{Singh2022, Morello2020}.}

{Another subtlety we noted is more detections for weaker signals (\snpp{} $<0.3$) at higher periods at $n_{\rm f} = 0.9$ (top four panels of Figure \ref{fig:fftnf}). At higher \snpp{}, this reverses to more detections at lower periods, which is the expected trend.}

{Initially, we processed the signals summing up to 8 harmonics (by default) for all period ranges; however, this introduced some inaccuracies and artefacts in the simulated results. As the candidate period is dependent on the highest harmonic summed, the candidate periods were not precise (e.g. the number of decimal places), which led to a decrement in detections (for $P = 1.5$\,s and $3$\,s), and some periods could not be detected ($P= 10$ and $50$\,s) at low \snpp. }
{When reprocessing the FFT search, summing up to 32 harmonics, we found a significant increase in the number of long-period detections and an increase in the precision of the detection period.}

\subsection{FFA}

RIPTIDE's \textbf{rffa} was very sensitive to all periods, signal strengths and nulling properties. It had the same if not more detections than \textbf{accelsearch} for each parameter set and was more consistent than \textbf{single\_pulse\_search} in detecting signals at multiple periods and of multiple strengths. Aside from the expected decrease in the number of detections for very high nulling properties at high periods, FFA was efficacious for a large range of parameters in comparison to FFT and SPS.

\subsection{SPS}

PRESTO's \textbf{single\_pulse\_search} was primarily only effective at high \snpp, as one would expect. Periodicity searches fold individual pulses to gain an integrated S/N, however, \textbf{single\_pulse\_search} does not benefit from that. At best, the algorithm can downsample the time series in order to increase the combined $\sigma_{\rm SPS}$. This is why \textbf{single\_pulse\_search} has very few detections until \snpp{}= 0.5.

\begin{figure*}[]
\vspace{-.85cm}
    \centering

    \begin{subfigure}[b]{.9\textwidth}
        \centering
        \caption{FFT}
        
        \begin{adjustbox}{trim=0 .95cm 0 .85cm,clip}{\includegraphics[width=\linewidth]{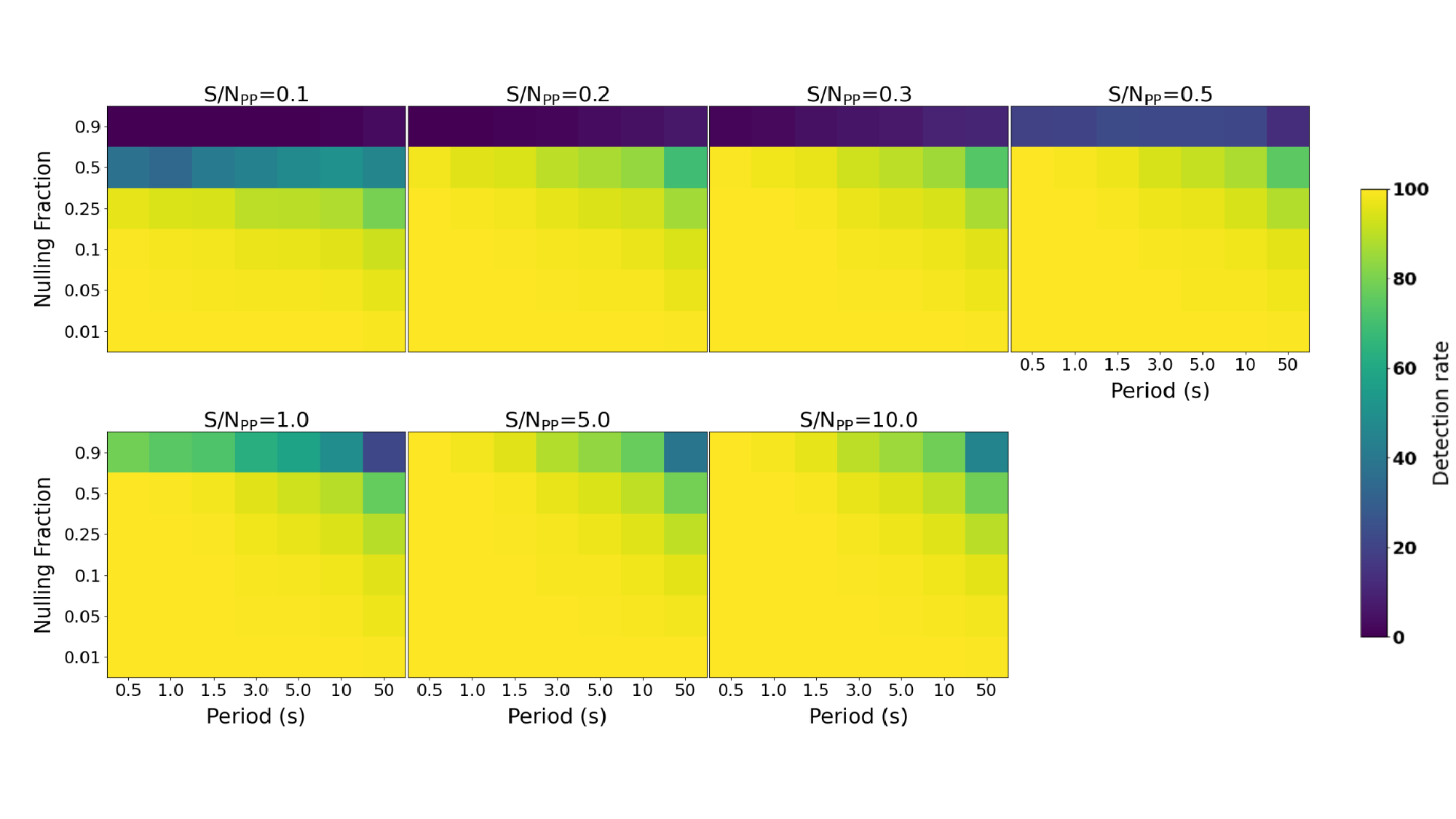}}
        \end{adjustbox}
        \label{fig:fftnf}
        
    \end{subfigure}
     
    \begin{subfigure}[b]{.9\textwidth}
         \centering
         \caption{FFA}
         
         \adjustbox{trim=0 .95cm 0 .85cm,clip}{\includegraphics[width=\linewidth]{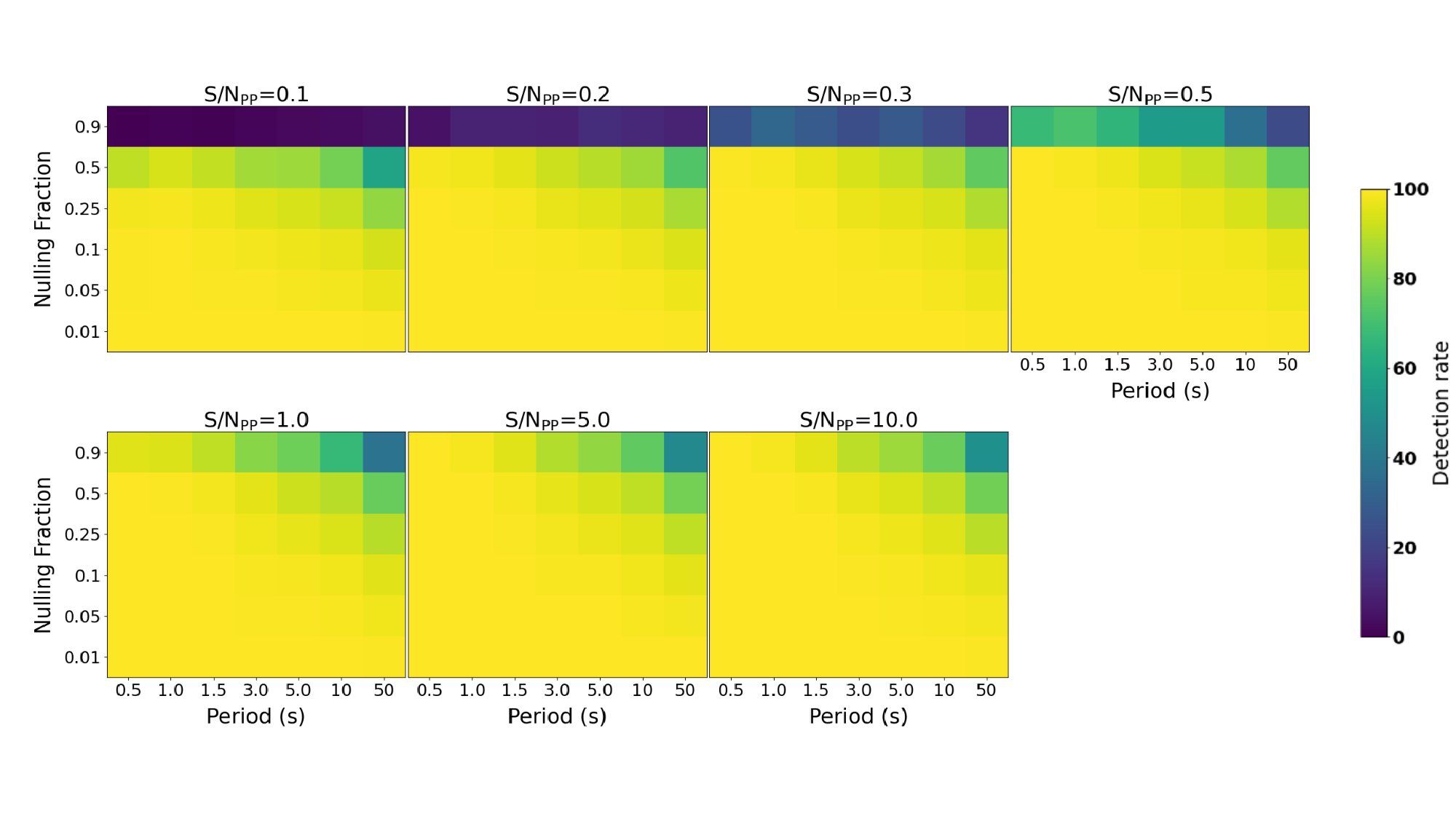}}
         \label{fig:ffanf}
         
    \end{subfigure}

    \begin{subfigure}[b]{.9\textwidth}
         \centering
         \caption{SPS}
         
         \adjustbox{trim=0 .95cm 0 .85cm,clip}{%
        \includegraphics[width=\linewidth]{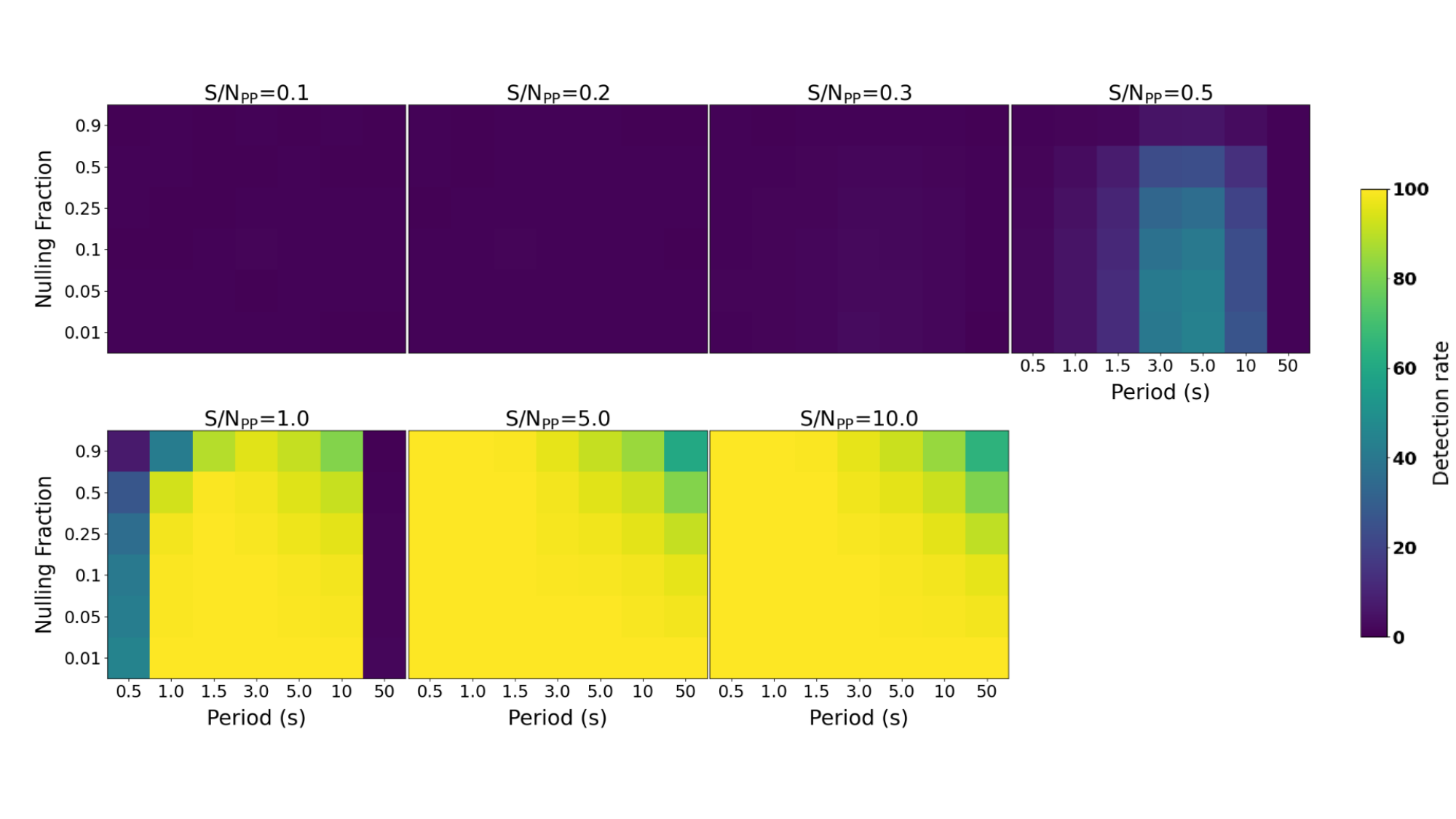}%
        }
         \label{fig:spsnf}
         
    \end{subfigure}
    \vspace{-.2cm}
    \caption{This figure shows the result of searching the simulated data with FFT (a), FFA (B) and SPS (c). The x-axis shows the period (s) and the y-axis shows the nulling fraction. The 7 plots inside the subfigures show heatmaps with the colour of each pixel indicating the percentage of signals, with the indicated period and nulling fraction, detected with the corresponding search method. This is accompanied by the colour bar on the right-hand side. The 7 plots represent different \snpp{} and are labelled as such.}
    \label{fig:pvnf}
\end{figure*}

The algorithm shows a clear preference for signals with periods between 1.5\,s - 10\,s as shown in the top right plots of Figures \ref{fig:spsnf} \& \ref{fig:SPSnd}. The \textbf{single\_pulse\_search} routine evaluates  $\sigma_{\rm SPS}$ of the pulse based on the matched filters of different widths and their convolution with the pulse. All pulses have the same duty cycle, meaning the pulse width scales with the period. Hence a matched filter would favour the larger period, which would have wider pulses. This is likely the reason why there are fewer detections for smaller periods at small \snpp{} for \textbf{single\_pulse\_search}. 

{However, there are a discrete number of matched filters with different widths, and as mentioned previously, the maximum widths of the filters by default range up to 30 bins. As the increment step ($dt$) of the time series is $3.2\times 10^{-3}$\,s, the width of the pulses for {the periods under consideration} go as 6.25, 12.5, 18.75, 37.5, 62.5, 125 and 625 bins. For $P=5$\,s where the pulse is $\sim$60 bins wide, which is double the maximum width allowed for the box car filter{, t}he filtering returns large $\sigma_{\rm SPS}$ which can allow for better detections; however, for larger periods, the pulse is too large and the maximum filter width of 30 is not able to encapsulate a significant enough portion of the pulse. This limits the $\sigma_{\rm SPS}$ and hence the detectability. Hence the feature we see is due to the choices of the filters and the shape of the pulses.}


At large \snpp{}, \textbf{single\_pulse\_search} performs effectively for both nulling parameters, and at times even performs better than \textbf{rffa}. Although it may be less meaningful to compare SPS with FFA or FFT as their methods have different criteria for detections, it is important to compare the efficacy of these methods to have the best likelihood of finding long-period and nulling pulsars.

\section{RESULTS FROM REAL DATA}\label{sec:realdata_results}

\subsection{Observational data}


We have compared the performance of various search methods on real survey data; we made use of example data from the ongoing SMART survey and measured the detection significance to assess the search methods. Specifically, we made use of data from the discovery and follow-up observations of new pulsars from the SMART survey. In all cases, the original detections were made using PRESTO's FFT implementation; the main purpose is to assess whether the inclusion of other search methods would have led to quicker or more significant detections of these pulsars. Observations of real pulsars used in this study are listed in Table \ref{tab:real_data}.

The observation lengths varied from 600\,s to 4800\,s, depending on the data that were readily available for analysis. Although all five pulsars have rotation periods of $\sim 1$\,s - 2\,s, two pulsars (\psrtwo \& \psrfive) also show significant amounts of nulling ($\sim$ 50\% and {33\%} respectively for PSRs \psrtwo \& \psrfive; cf. \citealt{2022ApJ...933..210M, Grover2024}, making them good targets for testing the efficacy of the methods explored in this work.

\begin{table*}[th]
\centering
\caption{The details of the observations and pulsars therein were used for the sanity check analysis. For more on these pulsars (excluding \psrfive) see \cite{Bhat2023smart2}. Analysis on \psrfive \citep{Grover2024}.}
    \begin{tabular}{c|c|c|c|c|c|c|c|c}
        
        Pulsar & Period (s) & DM \dmu & S/N$_{\rm FFA}$ & S/N$_{\rm FFT}$ & $\sigma_{\rm FFT}$ & $\sigma_{\rm SPS}$ & Observation length (s) & Observation MJD \\
        \hline
        \psrone	& 0.90 & 23.1(2) & 15.0&	22.2&	9.7&	5.3&	600& 58774\\
        \psrone	& 0.90 & 23.1(2) &17.5&	23.7&	10.6	&6.0	&2000 & 58792 \\
        \psrtwo	& 1.30 & 23.81(1) & 14.5&	21.5&	9.4&	9.2&	4800 & 58792\\
        \psrtwo	& 1.30 & 23.81(1) & 62.0&	54.6&	26.4&	14.0&	4800 & 58799\\
        \psrthree	& 1.67 & 42.28 &14.6&	18.9&	8.0&	5.7&	600 & 58890\\
        \psrfour	& 0.91 & 16.04 &11.5&	19.0	&8.2	&5.3	&900 & 59309 \\
        \psrfive	&1.66 & 19.77(2) &100.9	&118.2	&58.1	&10.1	&4200 & 58757 \\
        
    \end{tabular}
    \label{tab:real_data}
\end{table*}
\begin{figure*}
\vspace{-.85cm}
    \centering
    \begin{subfigure}[b]{0.9\textwidth}
    \caption{FFT}
    
        \centering
        \adjustbox{trim=0 .8cm 0 .75cm,clip}{\includegraphics[width=\linewidth]{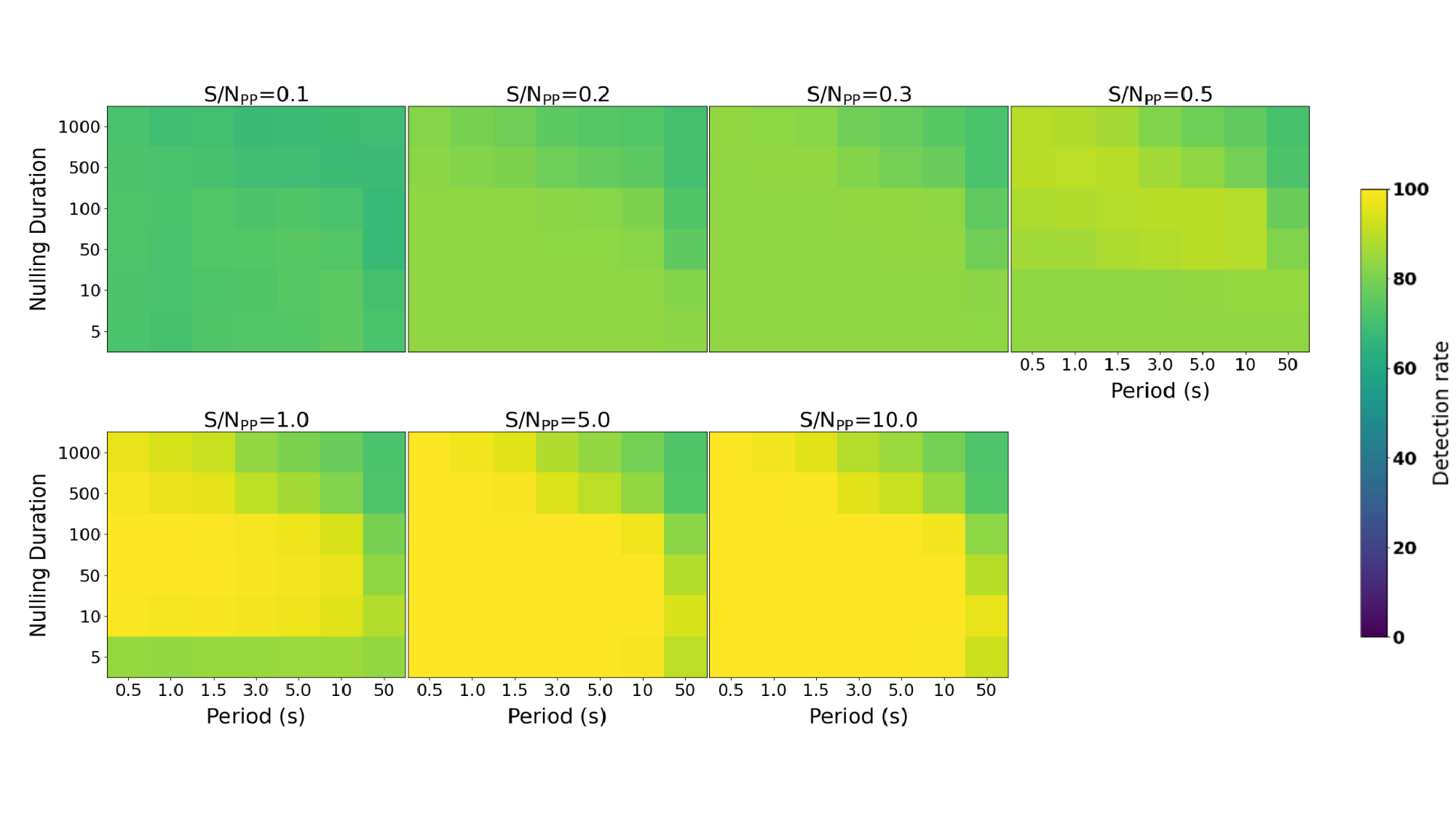}}
   
        \label{fig:FFTnd}
    \end{subfigure}

    \begin{subfigure}[b]{0.9\textwidth}
    \centering
    \caption{FFA}
  
    \adjustbox{trim=0 .8cm 0 .7cm,clip}{\includegraphics[width=\linewidth]{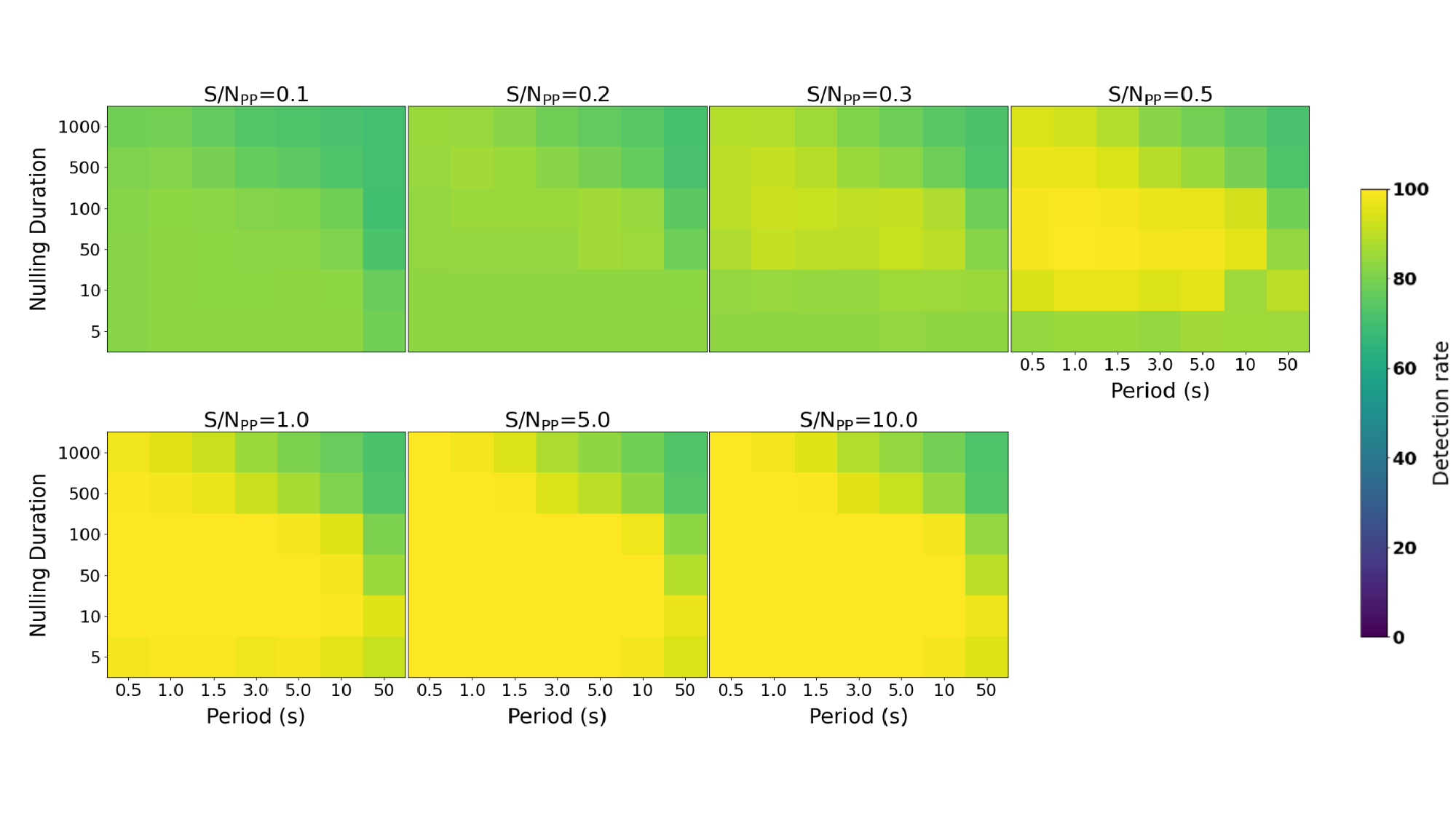}}

    \label{fig:FFAnd}
    \end{subfigure}

    \begin{subfigure}[b]{0.9\textwidth}
    \centering
    \caption{SPS}

    \adjustbox{trim=0 .9cm 0 .7cm,clip}{\includegraphics[width=\linewidth]{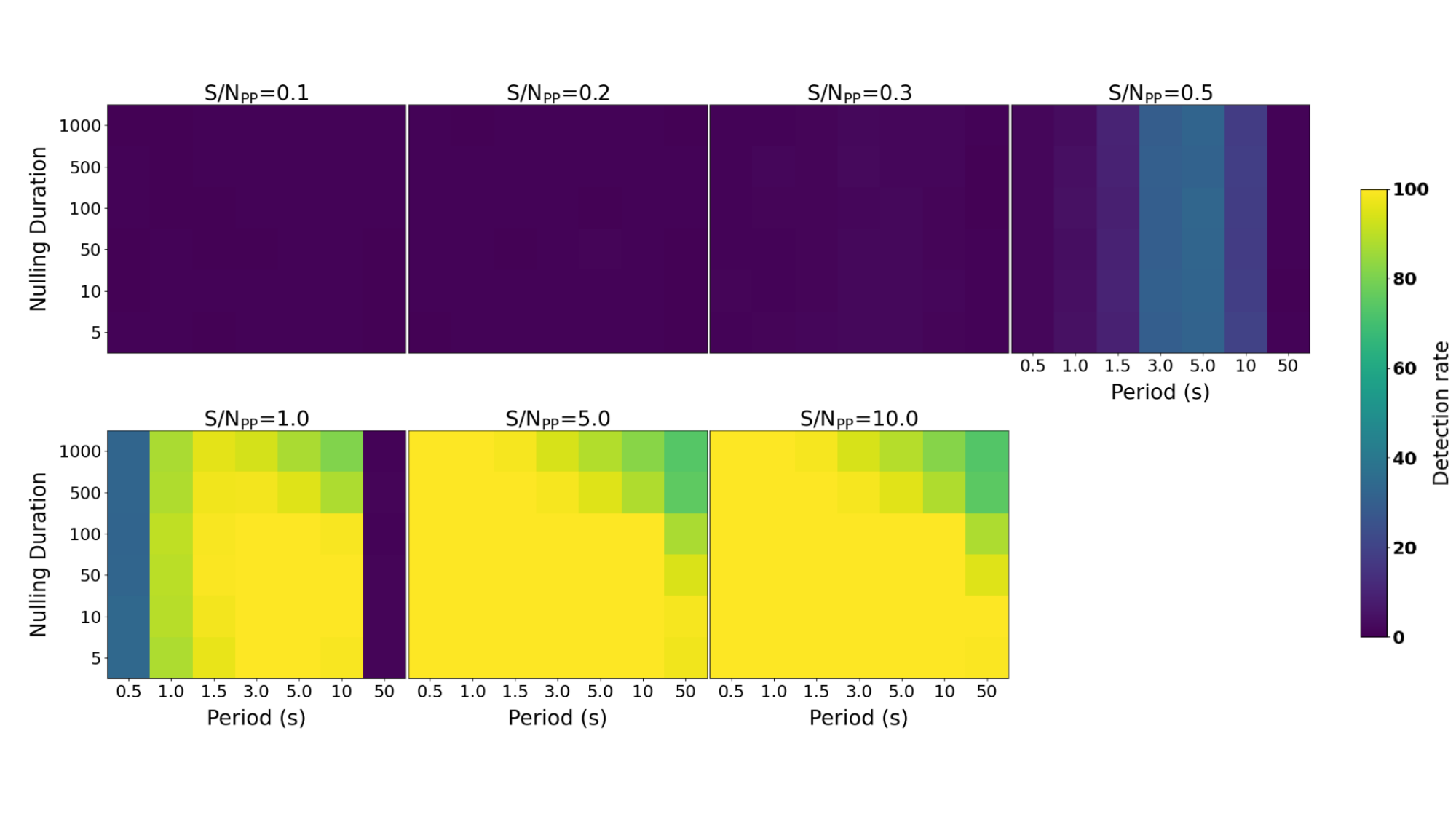}}
   
    \label{fig:SPSnd}
    \end{subfigure}

    \caption{This figure is very similar to Figure \ref{fig:pvnf}. It shows the result of searching the simulated data with FFT (a), FFA (B) and SPS (c). The x-axis shows the period (s) and the y-axis shows the nulling duration. The 7 plots inside the subfigures show heatmaps with the colour of each pixel indicating the percentage of signals, with the indicated period and nulling duration, detected with the corresponding search method. This is accompanied by the colour bar on the right-hand side. The 7 plots represent different \snpp{} and are labelled as such.}
    \label{fig:pnd}
\end{figure*}

\subsection{Quantitative performance}

 S22 compared the efficacies of FFA and FFT using S/N as their detection significance. They report RIPTIDE's S/N$_{\rm FFA}$ to be much larger than PRESTO's S/N$_{\rm FFT}$ for both simulated and real data. PRESTO's \textbf{accelsearch} reports a detection probability in the form of a different parameter, $\sigma_{\rm FFT}$, which may mistakenly be compared with \textbf{rffa}'s S/N$_{\rm FFA}$. Here, we have presented both values for a clear comparison (see Table \ref{tab:real_data}). 

We find S/N$_{\rm FFA}$ to be consistently larger than $\sigma_{\rm FFT}$, which is not meaningful since $\sigma_{\rm FFT}$ is essentially a probability statistic. However, we also find that S/N$_{\rm FFT}$ is larger in most cases than the S/N$_{\rm FFA}$, which is opposite to the findings of S22. 

SPS, on the other hand, provides a $\sigma_{\rm SPS}$, which is a multiple of the S/N$_{\rm raw}$. This value is consistently lower than $\sigma_{\rm FFT}$ value; however, the $\sigma_{\rm FFT}$ is based on many other parameters as well as the integrated profile, whereas the $\sigma$$_{\rm SPS}$ is based on individual pulses and the widths of the filters. Interestingly, we note that SPS returns a very high $\sigma$$_{\rm SPS}$ value for both nulling pulsars, with $\sigma$$_{\rm SPS}$ $\sim$ 9 and 10 for \psrtwo{} and \psrfive, respectively (as shown in Table \ref{tab:real_data}).

\section{DISCUSSION}\label{sec:dis}

On the basis of the analysis, we now comment on the efficacies of various search methods in detecting long-period and/or nulling pulsars. This is especially relevant for searches using wide-field instruments that also employ longer dwell times (e.g., the SMART and LOTAAS surveys). 

\subsection{Nulling duration vs Nulling fraction}


As described earlier (Section \ref{sec:meth}), nulling was characterised using two basic parameters: nulling fraction and nulling duration. Our results are shown in Figures \ref{fig:pvnf} \& \ref{fig:pnd}, where the detections (for signals with varying $n_{\rm f}$ and $N_{\rm d}$) are a function of the rotation period. At first glance, the number of detections for both parameters is similar for all search methods, particularly as they approach larger values of $n_{\rm f}$ and $N_{\rm d}$ at large \snpp.
It is difficult to compare $n_{\rm f}$ and $N_{\rm d}$ directly as the definition of 'large' is different for the two parameters. $n_{\rm f}$ has a clearly defined minimum and maximum value, whereas the range of $N_{\rm d}$ values are based entirely on observations.


Based on our analysis, the detections tend to vary more with an increase in $n_{\rm f}$ than that in $N_{\rm d}$, for the periodicity searches. At low \snpp{}, the number of detections is roughly constant for all values of $N_{\rm d}$. For larger \snpp{}, the detections tend to plateau near two to three values, where the number of detections remains consistent for some consecutive values of $N_{\rm d}$. This is only visible for FFT at higher \snpp{}, whereas for FFA, this is also observed at low \snpp. The plateauing is likely due to coarse increment sizes for $N_{\rm d}$.

The increments in $N_{\rm d}$ could be increased to better sample the range covered, allowing for a clear gradual change in the number of detections. In any case, the range we cover likely encapsulates the majority of nulling pulsars, as only a small percentage are known to null for longer durations, and as such, for larger values, we may expect to see only a small number of pulses from them. This is discussed further in Section \ref{sec:crit}.


\subsection{Comparison of the search methods on simulated data}


To analyse the efficacy of FFA, FFT and SPS implementations in detecting long-period and nulling pulsars,
{we used analysis of simulated data.}
As there were $1.764\times10^6$ individual time series streams to process, the computation time for the three search methods was essential to understand. This is a key factor that will influence the consideration of these methods for future surveys. Of all the three methods, the FFT implementation is the fastest and performs up to an order of magnitude faster compared to the other two. FFA implementation is the slowest. Although the results in Figures \ref{fig:ffanf} \& \ref{fig:FFAnd} show FFA to be more sensitive than FFT, FFT is computationally efficient. For long-duration surveys (e.g., the SMART survey, which can produce tens of millions of candidates), 
{the processing efficiency may outweigh the benefits}
of extra sensitivity.


The results presented in Section \ref{sec:simresults} suggest that FFA and FFT are surprisingly effective at finding pulsars with high nulling properties at low to moderate rotation periods. At larger periods and high nulling properties, both methods lose effectiveness as the signals become less periodic and less frequent. 

To further elaborate, \textbf{rffa} accurately and consistently detects pulsars throughout all \snpp. {PRESTO's \textbf{accelsearch} is also very effective, however, it has fewer detections at low signal strengths (\snpp{}=0.1 - 0.5) and high nulling fractions ($n_{\rm f}$ = 0.5 - 0.9). For longer periods ($P\geq 10$\,s), it heavily relies on the harmonics sums to be $\geq 16$ for the signal to be detected accurately at the fundamental frequency. } 

Our analysis also suggests that SPS is most effective in detecting long-period signals at high nulling {fractions and durations}, albeit this applies primarily at high \snpp{}. Although SPS may miss a large majority of generic and low-luminosity pulsars that FFT and FFA may detect, it is effective in detecting bright pulses, especially when emission is sporadic. This is vividly demonstrated by the way the extremely long-period pulsar, J0901-4046, was discovered, where the initial detection was made by searching for individual dispersed pulses \citep{Caleb2022}.

\subsection{Critique of the Simulation}\label{sec:crit}


Our choice of a detection significance above 6 was based on generic cut-offs for candidate selection typically employed in search pipelines. For SPS, a threshold of 6$\sigma$ was based on experiments with incorrectly de-dispersed data as described in Section 2.2.2. If the threshold was to be increased from 6 to 7 for example, it would primarily affect SPS. As FFA and FFT integrate the signal and base their significance values on the sum, their values are generally higher than 6, especially for simulated data in ideal noise. SPS's detection significance is based on the strength of individual pulses. Hence increasing the threshold from 6 to 7 would likely further decrease the detections made by SPS at low \snpp{}. However, by definition, this cut-off would also lead to more probable detections.

{The periods analysed in this study were spaced logarithmically to cover a large range of values. However, this led to large jumps in periods, specifically between 10\,s - 50\,s, which reflect as drastic changes in the detectability of the search methods in Figures \ref{fig:pvnf} \& \ref{fig:pnd}. Conducting a simulation for these intermediate periods would reveal more precisely how sensitive the methods are to detecting pulsars like J2251-3711 \citep{Tan2018} and J0901-4046 \citep{Caleb2022}.}




The limitation in terms of time and computational resources prompted us to consider simpler characteristics of pulsar signals; for example, we considered pulse shapes that are composed only of repeating Gaussian-shaped pulses with an additive Gaussian noise component to mimic receiver noise. This was sufficient for our purpose. Furthermore, as mentioned earlier in Section \ref{sec:meth}, the data were de-dispersed at a single DM of 150 \dmu. Exploring the impact of incorrect dedispersion (i.e. true DM different from an assumed DM) will involve more computation and resources, but can help to assess loss in detection significance resulting from an error in DM. 


Our analysis did not consider the effect of \emph{red noise} that may arise from variations in receiver noise or gain or varying levels of RFI.  As such, such effects tend to be specific to the characteristics of the instrument and therefore a generalisation may not be meaningful. In any case, we note that the inclusion of red noise would have affected the periodicity searches, making detections more difficult \citep{2017MNRAS.467.1661V}. We also note that S22 performed some useful comparison in this regard, between RIPTIDE and PRESTO, highlighting a significant reduction in the detection S/N in PRESTO searches than in RIPTIDE searches.

Furthermore, in our simulations, we allowed nulling to occur randomly (see Section \ref{sec:meth} for details). The choice of our considerations also meant that for some data sets, no pulses were present, due to either a high nulling fraction or a large nulling duration. In some ways, this reflects the realistic expectations in searching for objects with high degrees of intermittency. As a result, detections for \snpp{} = 5 and 10 look very similar for all three searches, and highlights the difficulty in finding such signals, regardless of their brightness. 

{The modulation was fixed to a value of 0.1, and we did not test how changing it might affect the different search methods. We can anticipate, however, that it would not have a strong effect on periodicity searches, which primarily rely on the integrated signal power and only to a lesser extent on the ``equivalent'' red noise introduced by the modulation.} However, a change in modulation would affect the SPS; e.g., a larger modulation index will allow certain pulses in the signal to be more readily detectable. I.e., intrinsically low \snpp{} signals would still have a higher probability of detection by SPS, especially when certain pulses are temporally boosted in brightness as a result of modulation. {This can be relevant for objects like giant pulse emitters and RRATs that tend to modulate heavily and emit bright pulses.}

The duty cycle of the pulse was also kept constant in our simulations. We note that previous studies (e.g. \citealt{Morello2020}, S22) have demonstrated the loss of sensitivity in FFT searches for narrow duty cycles.  We therefore chose to keep this constant.  Long-period pulsars tend to have very small duty cycles, $\lesssim 1\%$; however, a recently discovered pulsar-like source with a period of $\sim20$\,min showed a duty cycle of up to 25\%. For such small duty cycles, FFT's detection ability will be somewhat hindered, specifically for lower harmonics \citep{Morello2020}.

\subsection{Real data}

In order to verify the performance of various search methods, we tested \textbf{rffa}, \textbf{accelsearch} and \textbf{single\_pulse\_search} on examples of real data, chosen from SMART data sets that are currently being processed for a first-pass shallow survey (cf. \citealt{Bhat2023smart2}).

As summarised in Table \ref{tab:real_data}, both FFT and FFA comfortably found all five pulsars in all observations that we used. Based on the criteria employed for the simulated analysis, all sources would be classified as successful detections. We note that SPS  finds only two pulsars above a significance of 6; both nulling and bright enough to conduct single pulse analysis. This demonstrates that SPS is primarily sensitive to detecting bright pulsars.

In general, we find S/N$_{\rm FFT}$ to be higher than S/N$_{\rm FFA}$. The difference in S/N may arise due to the differences in the measurement of this value. For instance, FFA integrates the pulses in the time domain and measures the S/N from the combined profile, therefore the S/N$_{\rm FFA}$ is a function of the trial period and the trial width of the matched filters and is based on a defined Z-statistic \citep{Morello2020}. FFT instead sums the square root of the amplitude of the harmonics in the Fourier space to calculate S/N$_{\rm FFT}$ \citep{2001PhDT.......123R}. Theoretically, for an S/N measured in the Fourier domain to equal one in the time domain, the power in all harmonic bins (up to infinity) should be summed. This implies either RIPTIDE is underestimating the S/N or PRESTO is overestimating the S/N.

However, S22 report that S/N$_{\rm FFA}$ is up to orders of magnitude larger than S/N$_{\rm FFT}$ for a majority of their sources. A similar difference is seen in our case only when we compare S/N$_{\rm FFA}$ to $\sigma$$_{\rm FFT}$, the latter being the commonly adopted metric in \textbf{accelsearch} candidates.

\subsection{Boolean vs Quantitative comparison}

As mentioned earlier, comparing the different search methods by merely using commonly adopted detection significance {may be less meaningful than using a standard applicable to all methods.} Unravelling how the different values are connected is a complex task as there are many parameters and calculations involved. {This can lead to unfair comparisons between search methods as many generally assume a larger detection significance means a stronger detection; however, if the detection significance parameters are defined differently by two methods, then the larger value may not always mean a better detection.}

Further, this can also lead to confusion about which numbers to compare when conducting a thorough analysis. SPS lists $\sigma$$_{\rm SPS}$ as a detection significance of individual pulses; however,  FFT or FFA perform periodicity searches where they consider the sum of the pulses. Hence it is not meaningful to directly compare their S/N's and $\sigma$ values. Similarly, the $\sigma$$_{\rm FFT}$ should not be compared to S/N$_{\rm FFA}$ as they also measure different properties. This is made further difficult as the PRESTO implementation of FFT does not initially report a S/N$_{\rm FFT}$; it is obtained once candidates are pruned with \textbf{quick\_prune\_cands} or \textbf{ACCEL\_sift}.

Such inherent differences prompted us to consider a Boolean approach in our simulations where we ascertain the efficacies of the search methods based on their ability to make a detection. Although there may still be some bias due to the definition of the detection criteria, it is a more meaningful comparison for survey needs.



\subsection{Recommendations for future surveys}

Based on the performance of FFT, FFA and SPS on our simulated data, we outline some considerations to guide their effective use in long-duration surveys such as the SMART survey (see Table \ref{tab:recom}). 

Being a computationally efficient and inexpensive algorithm, FFT has been very effective for pulsar searching. Our analysis suggests that it is still the most effective method for searching pulsars with $P \lesssim 10$\,s. {Summing up to 32 harmonics, we found many long-period and nulling candidates, however, there was no red noise injected and the duty cycle of signals was kept constant, which can affect the effectiveness of FFT as noted by \cite{Singh2022, Morello2020}. FFT also detected fewer signals than FFA at low \snpp{} (0.1 - 0.5) and at high nulling fractions ($n_{\rm f} =0.5 - 0.9$). } Overall, the maximum period range that could be searched with FFT should be $P \lesssim 10$\,s. This accounts for the vast majority of the pulsar population. {This upper limit also means that the number of harmonics summed can be smaller (16 instead of 32), hence the processing can be faster.} 

FFA, in general, shows sensitivity throughout the parameter space; however, it is the slowest of the three search methods
{(by 1-2 order of magnitudes on local machine). }
{As FFT had fewer detections when summing weaker and highly nulling signals than FFA has been shown to be more sensitive to faint pulsars, which was also shown by \cite{Morello2020}, FFA should have some overlap with the FFT searching periods e.g. $P \gtrsim 1$\,s, to ensure a more thorough search.}
This is also computationally a less expensive option as \textbf{rffa} can be more than 6 times faster for $P \gtrsim 1$\,s \citep{Morello2020}.

SPS is also a computationally inexpensive algorithm that is independent of the exact period and can be quite effective in detecting sporadically emitting sources such as RRATs and FRBs. Most ongoing pulsar surveys routinely use this as part of their processing pipelines \citep[eg][]{2019A&A...626A.104S,2016ApJ...817..130B}, and as such it is among the future processing plans envisioned for SMART \citet{Bhat2023smart1}. However, we note it is primarily effective at detecting bright single pulses. This can be useful in detecting giant pulse emitters or bright RRATs. Considering the sporadicity aspect, SPS will offer increased sensitivity for detecting new objects, for longer observations. It appears that a higher detection threshold ($>$6 $\sigma_{\rm SPS}$) can be a more efficient way of finding
{bright sporadically emitting objects. }

Incorporation of the above considerations will help long-duration survey processing attain an increased detection sensitivity to a wider spectrum of pulsars and search a large part of the parameter space. This can be beneficial for future next-generation surveys with MeerKAT or SKA-Low, which have the potential to uncover the comprehensive population of radio-emitting neutron stars in our Galaxy. These can also be applicable to non-traditional search strategies such as those considered by CHIME
{where they take daily observations of the same position in the sky and sum their power spectra \citep[similar strategies to][]{2018ApJ...855..125C}. This allows for an essentially longer time series which can be useful in detecting long-period and nulling objects.}
Consideration of such extended search strategies will also enable ongoing and future searches to explore possible populations of pulsars that have been missed earlier due to selection effects.

\begin{table}[]
    \centering
    \begin{tabular}{c|c|c|c}
       Search method  & Software & Period range & $\sigma$ or S/N threshold\\
       \hline
    FFT & PRESTO  & $\lesssim$ 10\,s & $\sim$ 6 \\
        FFA &RIPTIDE & $\gtrsim$ {1}\,s& $\sim$ 6 \\
       SPS &PRESTO  & - & $ > $ 6
    \end{tabular}
    \caption{A summary of the recommendations for the three search algorithms based on this work.}
    \label{tab:recom}
\end{table}


    

\section{CONCLUSION}
\label{sec:con}

To develop a more comprehensive picture of pulsar emission properties and various aspects relating to their populations,  it is important to extend current search approaches to include algorithms and techniques that will allow a more extensive exploration of the long-period parts of the parameter space. This can potentially lead to an improved understanding of pulsar death lines and the related physical modes, alongside yielding new insights into linking nulling to long periods, as hinted by some of the recent discoveries. With the advent of wide-field radio-astronomy facilities like the MWA and LOFAR, and their adoption for large (untargeted) pulsar surveys, longer dwell times per pointing become more affordable, and this can be well exploited for increasing the detection sensitivity to such long-period and nulling pulsars. Popular search algorithms and software tend to vary both in their efficacy and sensitivity to the related classes of objects. Motivated by this, we explored the efficacies of different search methods in detecting long-period and nulling pulsars, using both simulations and examples of real data. The software used includes PRESTO's FFT implementation (\textbf{accelsearch}), RIPTIDE's FFA implementation (\textbf{rffa}) and PRESTO's SPS implementation (\textbf{single\_pulse\_search}). The simulation analysis was approached from the perspective of exploring a wider range in parameter space.

Our analysis finds FFA to be most sensitive to pulsar signals over a much wider range in period ($\sim$ tens of seconds) even when the signal strengths are low;  however, it is also slower than FFT. {While FFT is computationally more efficient and generally sensitive towards long periods, it loses sensitivity for very weak and highly nulling signals }. SPS on the other hand is primarily sensitive to sporadic or intermittent emission, but largely sensitive to high signal strengths. Therefore, a more effective (if not ideal) search strategy could include the use of FFT searches for $P \lesssim 10$\,s, FFA searches for {$P \gtrsim 1$\,s}, alongside  SPS with a detection threshold of $ > 6 \sigma$, and preferably with {long observations ($>10$ minutes)}, for increased sensitivity to intermittent or sporadic emitters.  While 
{such a strategy was derived from }
simulations and examples of real data selected from SMART survey data sets, they are also applicable for many of the current and planned large pulsar surveys, including those envisioned with SKA-Low. In general, more extensive explorations of the long-period parts of the search parameter space can help overcome some of the inherent observational biases in detecting long-period or nulling pulsars, and thus uncover a larger population (and variety) of objects that will help us to develop a better understanding of pulsar population and emission physics. 

\begin{acknowledgement}

The authors would like to acknowledge Scott Ransom and Vincent Morello for discussions and advice on optimal ways to run the search software. Bradley Meyers and Chia Min Tan for their help in the analysis and review process, and Edwin Dickinson for their useful comments.

This scientific work uses data obtained from Inyarrimanha Ilgari Bundara / the Murchison Radio-astronomy Observatory. We acknowledge the Wajarri Yamaji People as the Traditional Owners and native title holders of the Observatory site. Establishment of CSIRO's Murchison Radio-astronomy Observatory is an initiative of the Australian Government, with support from the Government of Western Australia and the Science and Industry Endowment Fund. Support for the operation of the MWA is provided by the Australian Government (NCRIS), under a contract to Curtin University administered by Astronomy Australia Limited. This work was supported by resources provided by the Pawsey Supercomputing Research Centre with funding from the Australian Government and the Government of Western Australia.

This work was supported by resources awarded under Astronomy Australia Ltd's ASTAC merit allocation scheme on the OzSTAR national facility at Swinburne University of Technology. The OzSTAR program receives funding in part from the Astronomy National Collaborative Research Infrastructure Strategy (NCRIS) allocation provided by the Australian Government, and from the Victorian Higher Education State Investment Fund (VHESIF) provided by the Victorian Government.

\end{acknowledgement}

\bibliography{example}



\end{document}